\documentclass[aps,prl,reprint,groupedaddress,nofootinbib,superscriptaddress]{revtex4-1}

\usepackage[T1]{fontenc}
\usepackage[utf8]{inputenc}

\usepackage{amsmath} 
\usepackage[normalem]{ulem}
\usepackage{xcolor}
\usepackage{graphicx}
\usepackage{ulem}
\usepackage{multirow}
\usepackage{bm}
\graphicspath{ {figures/} }

\begin{document}

\title{Tests of General Relativity with GW170817}


\author{B.~P.~Abbott}
\affiliation{LIGO, California Institute of Technology, Pasadena, CA 91125, USA}
\author{R.~Abbott}
\affiliation{LIGO, California Institute of Technology, Pasadena, CA 91125, USA}
\author{T.~D.~Abbott}
\affiliation{Louisiana State University, Baton Rouge, LA 70803, USA}
\author{F.~Acernese}
\affiliation{Universit\`a di Salerno, Fisciano, I-84084 Salerno, Italy}
\affiliation{INFN, Sezione di Napoli, Complesso Universitario di Monte S.Angelo, I-80126 Napoli, Italy}
\author{K.~Ackley}
\affiliation{OzGrav, School of Physics \& Astronomy, Monash University, Clayton 3800, Victoria, Australia}
\author{C.~Adams}
\affiliation{LIGO Livingston Observatory, Livingston, LA 70754, USA}
\author{T.~Adams}
\affiliation{Laboratoire d'Annecy de Physique des Particules (LAPP), Univ. Grenoble Alpes, Universit\'e Savoie Mont Blanc, CNRS/IN2P3, F-74941 Annecy, France}
\author{P.~Addesso}
\affiliation{University of Sannio at Benevento, I-82100 Benevento, Italy and INFN, Sezione di Napoli, I-80100 Napoli, Italy}
\author{R.~X.~Adhikari}
\affiliation{LIGO, California Institute of Technology, Pasadena, CA 91125, USA}
\author{V.~B.~Adya}
\affiliation{Max Planck Institute for Gravitational Physics (Albert Einstein Institute), D-30167 Hannover, Germany}
\affiliation{Leibniz Universit\"at Hannover, D-30167 Hannover, Germany}
% ...

\collaboration{The LIGO Scientific Collaboration and the Virgo Collaboration}



\date{\today}

\begin{abstract}
The recent discovery by Advanced LIGO and Advanced Virgo of a gravitational wave signal from a binary neutron star inspiral has enabled tests of general relativity (GR) with this new type of source.
This source,  for the first time, permits tests of strong-field dynamics of compact binaries in presence of matter. 
In this paper, we place constraints on the dipole radiation and possible deviations from GR in the post-Newtonian coefficients that govern the inspiral regime.
Bounds on modified dispersion of gravitational waves are obtained; in combination with information from the observed electromagnetic counterpart we can also constrain effects due to large extra dimensions.
Finally, the polarization content of the gravitational wave signal is studied.
The results of all tests performed here show good agreement with GR.

\end{abstract}


\maketitle

\section{Introduction}

On August 17, 2017 at 12:41:04 UTC, the Advanced LIGO and Advanced Virgo gravitational-wave (GW) detectors made their first observation of a binary neutron star inspiral signal, called GW170817 \cite{TheLIGOScientific:2017qsa}.
Associated with this event, a gamma ray burst \cite{Monitor:2017mdv} was independently observed, and an optical counterpart was later discovered \cite{GBM:2017lvd}.
In terms of fundamental physics, these coincident observations led to a stringent constraint on the difference between the speed of gravity and the speed of light; allowed new bounds to be placed on local Lorentz invariance violations; and enabled a new test of the equivalence principle by constraining the Shapiro delay between gravitational and electromagnetic radiation \cite{Monitor:2017mdv}.
These bounds, in turn, helped to strongly constrain the allowed parameter space of alternative theories of gravity that offered gravitational explanations for the origin of dark energy \cite{Baker:2017hug,Creminelli:2017sry,Sakstein:2017xjx,Ezquiaga:2017ekz,Langlois:2017dyl,Dima:2017pwp,deRham:2018red} or dark matter \cite{Boran:2017rdn}.

In this paper we present a range of tests of general relativity (GR) that have not yet been done with GW170817.
Some of these are extensions of tests performed with previously
discovered binary black hole coalescences
\cite{Abbott:2016blz,TheLIGOScientific:2016src,Abbott:2016nmj,TheLIGOScientific:2016pea,Abbott:2017vtc,Abbott:2017oio,Abbott:2017gyy},
an important difference being that the neutron stars' tidal
deformabilities need to be taken into account in the waveform models.
The parameter estimation settings for this analysis broadly
match with those of \cite{Abbott:2018wiz,Abbott:2018exr} which reported the
properties of the source GW170817. 
Our approach here is theory-agnostic where, using GW170817, we constrain
generic features in the gravitational waveform that may
arise from a breakdown of GR. For
a detailed discussion about specific alternative theories
that predict one or more of the physical effects discussed
here see, for instance, Sec.~5 of ~\cite{Gair:2012nm} and Sec.~2 of \cite{Yunes:2013dva}.

Three types of tests are presented.
In Sec.~II, we study the general-relativistic dynamics of the source, in particular constraining
dipole radiation in the strong-field and radiative regime and checking for possible deviations in 
the post-Newtonian (PN) description of binary inspiral by studying the phase evolution of the signal.
Sec.~III focuses on the way gravitational waves propagate over large distances.
Here we look for anomalous dispersion, which enables complementary bounds on violations of 
local Lorentz invariance to those of \cite{Monitor:2017mdv}; constraints on large extra spatial dimensions are obtained by comparing the distance inferred from the GW signal with the one inferred from the electromagnetic counterpart.
Finally, in Sec.~IV constraints are placed on alternative polarization states, where this time the position of the source on the sky can be used, again because of the availability of an electromagnetic counterpart.
We end with a summary and conclusions.

\section{Constraints on deviations from the general-relativistic dynamics of the source}

\begin{figure*}
\includegraphics[width=\textwidth]{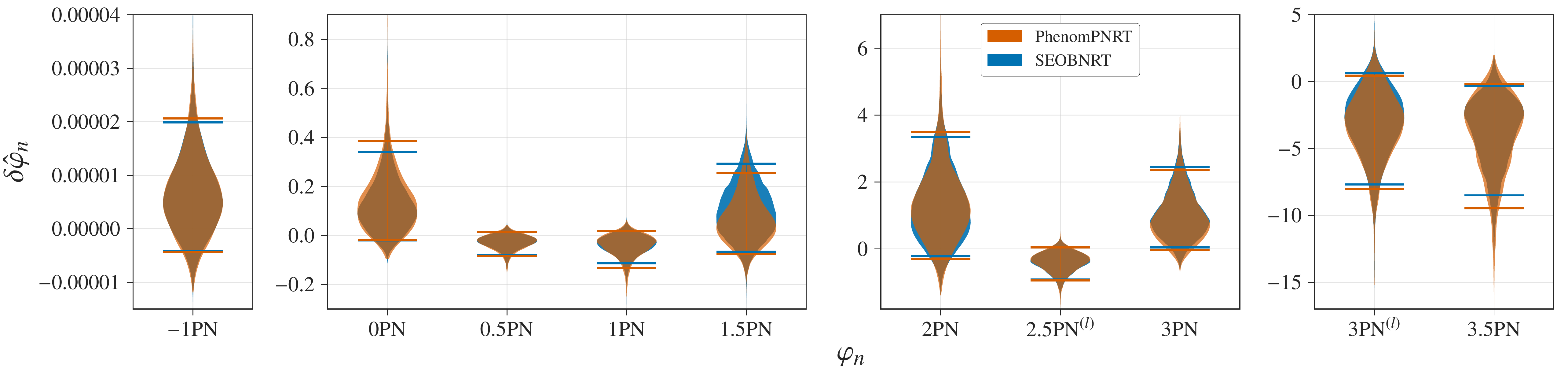}
\caption{Posterior density functions on deviations of PN coefficients $\delta\hat\varphi_n$ obtained using two different waveform models (\texttt{PhenomPNRT} and \texttt{SEOBNRT}); see the main text for details.
The $-1$PN and $0.$5PN corrections correspond to absolute deviations, whereas all others represent fractional deviations from the PN coefficient in GR. 
The horizontal bars indicate 90\% credible regions.}
\label{fig:violinplots}
\end{figure*}

\begin{figure}
\includegraphics[width=0.9\columnwidth]{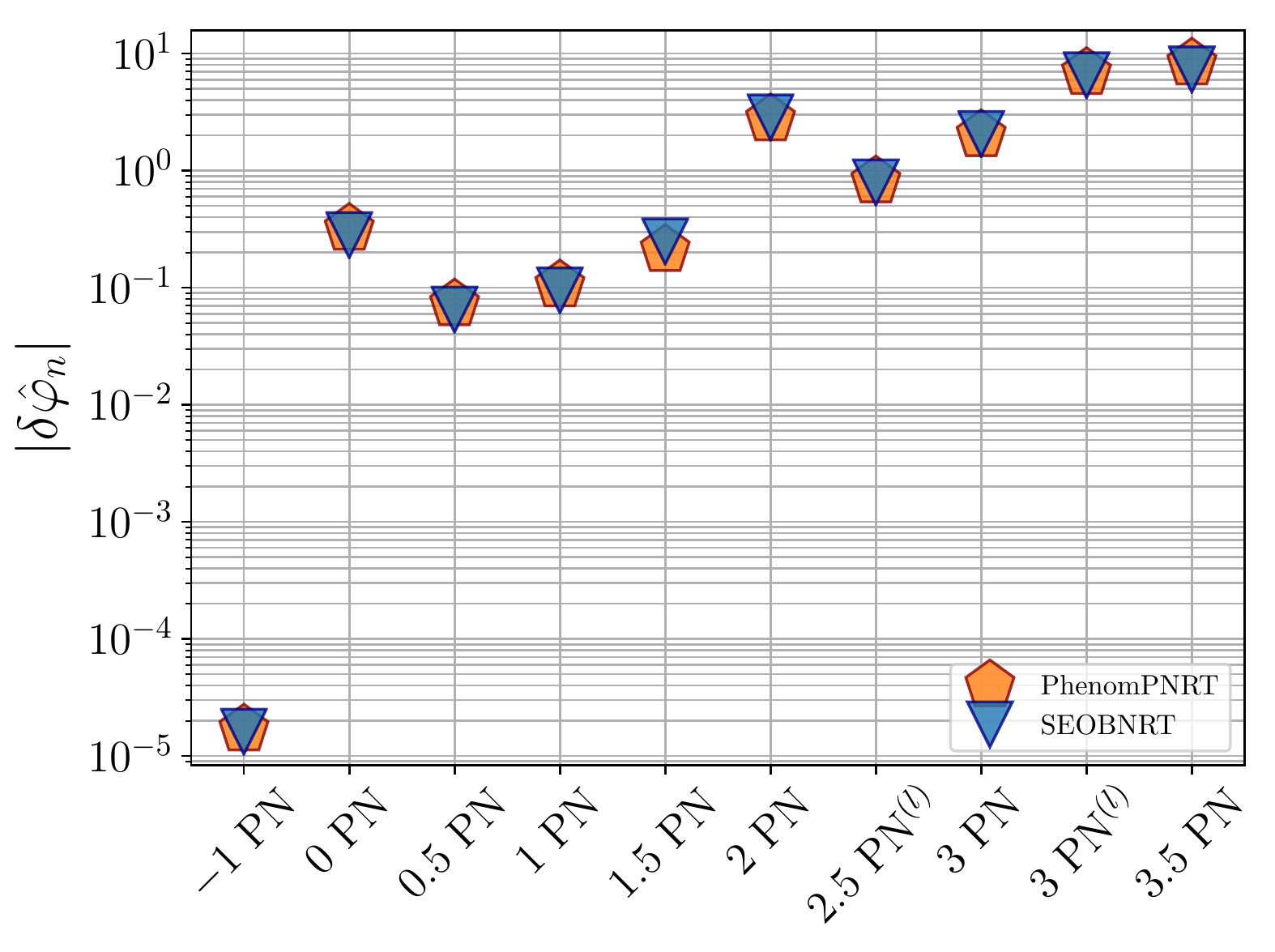}
\caption{90\% upper bounds on deviations $|\delta\hat\varphi_n|$ in the PN coefficients
following from the posterior density functions shown in Fig.~\ref{fig:violinplots}.}
\label{fig:pnbounds}
\end{figure}

Testing GR via the dynamics of a binary system involves constructing a waveform model that allows for parameterized deformations away from the predictions of GR and then constraining the associated parameters that govern those deviations \cite{Blanchet:1994ez,Arun:2006hn,Yunes:2009ke,Mishra:2010tp,Li:2011cg,Meidam:2017dgf,TheLIGOScientific:2016src,TheLIGOScientific:2016pea,Abbott:2017vtc}.
For previous observations of coalescing binary black
holes~\cite{TheLIGOScientific:2016src,TheLIGOScientific:2016pea}, these
tests relied on the frequency domain \texttt{IMRPhenomPv2} waveform
model of \cite{Husa:2015iqa,Khan:2015jqa,Hannam:2013oca}, which
describes the inspiral, merger, and ringdown of vacuum black holes, and
provides an effective description of spin precession making
the best use of the results from analytical and numerical
relativity~\cite{Blanchet:1995ez,Blanchet:2004ek,Kidder:1995zr,Blanchet:2006gy,Blanchet:2011zv,Mikoczi:2005dn,Bohe:2015ana,Marsat:2014xea}.
The phase evolution of this waveform is governed by a set of coefficients $p_n$ that depend on the component masses and spins. 
These coefficients include post-Newtonian (PN) parameters and phenomenological constants that are calibrated against numerical relativity waveforms to describe the intermediate regime between inspiral and merger, as well as the merger-ringdown.
To test GR, the waveform model is generalized to allow for relative deviations in each of the coefficients 
in turn, \emph{i.e.}~by replacing $p_n \rightarrow (1 + \delta\hat{p}_n)\,p_n$, where the $\delta\hat{p}_n$ 
are zero in GR.
The $\delta\hat{p}_n$ are then varied along with all the parameters that are also present in the case of GR (masses, spins, and extrinsic parameters), and posterior density functions (PDFs) 
are obtained using {\tt LALInference}~\cite{Veitch:2014wba}.
For GR to be correct, the value $\delta\hat{p}_n = 0$ should fall within the support of each of the PDFs.
Note that although one could also let \emph{all} of the testing parameters vary at the same time, 
this will tend to lead to uninformative posteriors (see e.g.~\cite{TheLIGOScientific:2016src}). Fortunately, 
as demonstrated in Refs.~\cite{Sampson:2013lpa,Li:2011cg,Li:2011vx,Meidam:2017dgf}, checking for a deviation 
from zero in a single testing parameter is an efficient way to uncover GR violations that occur 
at multiple PN orders, and one can even find violations at powers of frequency that are distinct 
from the one that the testing parameter is associated with \cite{Li:2011cg,Li:2011vx,Meidam:2017dgf}. 
Thus, such analyses are well suited to search for generic departures from GR. There is 
a limitation though: Should a GR 
violation be present, then the measured values of the $\delta\hat{p}_i$ will not necessarily reflect 
the predictions of whichever alternative theory happens to be the correct one.   
To reliably constrain theory-specific quantities such as coupling constants or extra charges, one should 
utilize full inspiral waveform models from specific modified gravity theories, including modifications
to the phase at all the orders where they appear. Unfortunately, in most cases only leading-order 
modifications are available at the present time \cite{2016PhRvD..94h4002Y}. However, in this section 
the focus is mostly on model-independent tests of general relativity itself.

In this work, we modify this approach in two ways.
First, we use waveform models more suitable for binary neutron stars.
Second, whereas the infrastructure~\cite{Li:2011cg} used to test GR with binary black holes observations~\cite{TheLIGOScientific:2016src,TheLIGOScientific:2016pea} was restricted to waveform models that depend directly on the coefficients $p_n$, we also introduce a new procedure that can include deviations to the phase evolution parameterized by $\delta \hat{p}_n$ to any frequency domain waveform model.
We conduct independent tests of GR using inspiral-merger-ringdown models that incorporate deviations from GR using each of these two prescriptions; by comparing these analyses, we are able to estimate the magnitude of systematic modeling uncertainty in our results.

The merger and ringdown regimes of binary neutron stars differ from those of binary black holes, and tidal effects not present in binary black holes need to be included in the description of the inspiral.
Significant work has been done to understand and model the dynamics of binary neutron stars analytically using the PN approximation to general relativity~\cite{Blanchet:2013haa}.
This includes modeling the
non-spinning~\cite{Blanchet:1995ez,Blanchet:2004ek} and spinning
radiative (or inspiral) dynamics~\cite{Kidder:1995zr,Blanchet:2006gy,Blanchet:2011zv,Mikoczi:2005dn,Bohe:2015ana,Marsat:2014xea} as well as finite size effects~\cite{Flanagan:2007ix,Hinderer:2009ca,Vines:2011ud} for binary neutron star systems.
Frequency domain waveforms based on the stationary phase approximation~\cite{Sathyaprakash:1991mt} have been developed incorporating the abovementioned effects~\cite{Buonanno:2009zt,Arun:2008kb,Mishra:2016whh} and have been successfully employed for the data analysis of compact binaries.
A combination of these analytical results with the results from numerical relativity simulations of binary neutron star mergers (see \cite{Faber:2012rw} for a review) have led to the development of efficient waveform models which account for tidal effects~\cite{Bernuzzi:2014owa,Hinderer:2016eia,Dietrich:2017aum}.

We employ the \texttt{NRTidal} models introduced in~\cite{Dietrich:2017aum,Dietrich:2018uni} as the basis 
of our binary neutron star waveforms: frequency domain waveform models for binary black holes are 
converted into waveforms for inspiraling neutron stars that undergo tidal deformations by adding to 
the phase an appropriate expression $\phi_T(f)$ and windowing the amplitude such that the merger and 
ringdown are smoothly removed from the model; see \cite{Dietrich:2018uni} for details.
The closed-form expression for $\phi_T(f)$ is built by combining PN information, the tidal effective-one-body (EOB) model of~\cite{Bernuzzi:2014owa}, and input from numerical relativity (NR).
We consider two waveform models that use this description of tidal effects.
 One of these models---\texttt{IMRPhenomPNRT}, detailed below---describes a binary neutron star with
precessing spins. Though the form of $\phi_T(f)$ was originally obtained in a setting where the
neutron stars were irrotational or had their spins aligned to the angular momentum, tides can be
included in this waveform model by first applying $\phi_T(f)$ to an aligned-spin waveform, and then
performing the twisting-up procedure that introduces spin precession~\cite{Schmidt:2012rh}.

The first binary neutron star model we consider is constructed by applying this procedure to \texttt{IMRPhenomPv2} waveforms.
Following the nomenclature of~\cite{Abbott:2018wiz}, we refer to the resulting waveform model as \texttt{PhenomPNRT}.
Parameterized deformations $\delta\hat{p}_n$ are then introduced as shifts in parameters describing the phase in precisely the same way as was done for binary black holes.
This will allow us to naturally combine PDFs for the $\delta\hat{p}_n$ from measurements on binary black holes and binary neutron stars, arriving at increasingly sharper results in the future.
Because of the unknown merger-ringdown behavior in the case of binary neutron stars, which in any case gets removed from the waveform model, in practice only deviations $\delta\hat{\varphi}_n$ in the PN parameters $\varphi_n$ can be bounded.
The set of possible testing parameters is taken to be
\begin{equation}
\{ \delta\hat\varphi_{-2}, \delta\hat\varphi_0, \delta\hat\varphi_1, \delta\hat\varphi_2, \delta\hat\varphi_3, \delta\hat\varphi_4, \delta\hat\varphi_5^{(\ell)}, \delta\hat\varphi_6, \delta\hat\varphi_6^{(\ell)}, \delta\hat\varphi_7 \},
\end{equation}  
where the $\delta\hat\varphi_n$ are associated with powers of frequency $f^{(-5+n)/3}$, and $\delta\hat\varphi_5^{(\ell)}$ and $\delta\hat\varphi_6^{(\ell)}$ with functions $\log(f)$ and $\log(f)\,f^{1/3}$, respectively; 
$\delta\hat\varphi_5$ would be completely degenerate with some reference phase $\phi_c$ and hence is not included in the list.
In addition to corrections to the positive PN order coefficients, deviations at $-1$PN are included because they offer the possibility to constrain the 
presence of dipole radiation during the inspiral (discussed below).  
$\delta\hat\varphi_{-2}$ and $\delta\hat\varphi_1$ represent absolute rather than relative 
deviations, as both are identically zero in GR.

We also employ the \texttt{SEOBNRv4} waveform model, which is constructed from an aligned-spin EOB model for binary black holes augmented with information from NR simulations~\cite{Bohe:2016gbl}.
Using the methods of~\cite{Purrer:2014fza}, this model is evaluated in the frequency domain, and then we add the tidal correction $\phi_T(f)$ as described above; we refer to the resulting waveform model as \texttt{SEOBNRT}.
Unlike \texttt{PhenomPNRT}, the \texttt{SEOBNRT} model is not constructed explicitly in terms of PN coefficients $\varphi_n$.
Instead, we model the effect of a relative shift $\delta\hat\varphi_n$ by adding to the frequency domain phase a term $\delta\hat\varphi_n\,\varphi_n\,f^{(-5+n)/3}$ or $\delta\hat\varphi_n^{(\ell)}\,\varphi_n^{(\ell)}\,f^{(-5+n)/3}\,\log(f)$, as applicable.
These corrections are then tapered to zero at the merger frequency.

Fig.~\ref{fig:violinplots} depicts the PDFs on $\delta \hat \varphi_n$ recovered when only variations at that particular PN order are allowed.
We find that the phase evolution of GW170817 is consistent with the GR prediction.
The 90\% credible region for each parameter contains the GR value of $\delta\hat\varphi_n = 0$ at 
all orders other than 3PN and 3.5PN.\footnote{Using \texttt{PhenomPNRT} (\texttt{SEOBNRT}), 
the GR value lies at the 6.8-th (4.4-th) percentile of the PDF for the 3PN parameter and at the 95.0-th 
(96.7-th) percentile for the 3.5PN parameter.} For the pipeline used to perform parameterized tests
with binary black holes, it has been shown in \cite{Meidam:2017dgf} through extensive simulations that 
no noticeable systematics are present. In the case of binary neutron stars such a study is computationally 
demanding because of the long signals, and a similar study will be published at a later date. 
At present we have no reason to believe that the offsets seen here at 3PN and 3.5PN have anything 
other than a statistical origin. In any case, we note that the value of zero is in the support of 
the posterior density function for all testing parameters.
The bounds on the positive-PN parameters ($n \geq 0$) obtained with GW170817 alone are comparable to those obtained by combining the binary black hole signals GW150914, GW151226, and GW170104 in \cite{Abbott:2017vtc} using the \texttt{IMRPhenomPv2} waveform model.
For convenience we also separately give 90\% upper bounds on deviations in PN coefficients; see 
Fig.~\ref{fig:pnbounds}.

The PDFs shown in Fig.~\ref{fig:violinplots} were constructed using the same choice of prior distribution outlined in~\cite{Abbott:2018wiz} with the following modifications.
We use uniform priors on $\delta \hat{\varphi}_n$ that are broad enough to fully contain the plotted PDFs. 
Due to the degeneracy between $\delta \hat{\varphi}_0$ and the chirp mass, a broader prior distribution was chosen for the latter as compared to in~\cite{Abbott:2018wiz} for runs in which $\delta \hat{\varphi}_0$ was allowed to vary.
All inference was done assuming the prior $|\bm{\chi}_i|\leq 0.99$, where $\bm{\chi}_i = c \mathbf{S}_i / (G m_i^2)$ is the dimensionless spin of each body. 
This conservative spin prior was chosen to allow the constraints on $\delta \hat{\varphi}_n$ to be directly compared with those from binary black hole observations, which used the same prior~\cite{TheLIGOScientific:2016src,TheLIGOScientific:2016pea}.
Nevertheless, throughout this paper we assume the two objects to be neutron stars, 
and following \cite{Abbott:2018wiz} we limit our prior on the component tidal parameters 
to $\Lambda_i \leq 5000$. (For a precise definition of the $\Lambda_i$, see
\cite{TheLIGOScientific:2017qsa} and references therein.) This choice was motivated by reasonable 
astrophysical assumptions 
regarding the expected ranges for neutron star masses and equations of 
state~\cite{Hinderer:2009ca,Lattimer:2012nd,Suwa:2018uni}; higher values of $\Lambda$ are possible for 
some equations of state if the neutron star masses are small ($\simeq 0.9\,M_\odot$). 
The extra freedom introduced by including $\delta \hat{\varphi}_n$ leads to a loss in 
sensitivity in the measurement of tidal parameters; in particular, the tail of the 
PDF for the tidal deformation of the less massive body $\Lambda_2$ touches the prior 
upper bound in many of the tests.
The correlation between $\delta \hat{\varphi}_n$ and $\Lambda_2$ means 
that the upper bounds for $|\delta \hat{\varphi}_n|$ would be weaker if we did 
not impose our neutron star prior of $\Lambda_i \leq 5000$.

Certain differences are present between the \texttt{PhenomPNRT} and \texttt{SEOBNRT} waveform models and the way they are used.
First, \texttt{PhenomPNRT} allows for precessing spin configurations, whereas the \texttt{SEOBNRT} is restricted to systems with spins aligned with the orbital angular momentum.
Second, continuity conditions enforced in the construction of \texttt{PhenomPNRT} waveforms cause deviations from GR in the inspiral to affect the behavior of later phases of the signal, whereas the tapering of deviations in \texttt{SEOBNRT} ensures that the merger-ringdown of the underlying waveform is exactly reproduced.
However, this discrepancy is not expected to affect measurements of $\delta\hat\varphi_n$ significantly, because the signal is dominated by the inspiral, and both waveform models are amplitude-tapered near merger.
Third, the spin-induced quadrupole moment \cite{Poisson:1997ha}, which enters the phase at $2$PN through quadrupole-monopole couplings, is computed using neutron-star universal relations~\cite{Yagi:2013bca} in \texttt{PhenomPNRT} and is assumed to take the black-hole value in \texttt{SEOBNRT}.
Finally, in the \texttt{PhenomPNRT} model, fractional deviations are applied only to non-spinning terms in the PN expansion of the phase, \emph{i.e.}~terms dependent on the bodies' spins retain their GR 
values\footnote{There is no fundamental reason for this choice; we follow the convention used 
in previous publications on parameterized tests of GR \cite{Li:2011cg,TheLIGOScientific:2016src,TheLIGOScientific:2016pea,Abbott:2017vtc,LIGOScientific:2019fpa}.}, while in \texttt{SEOBNRT}, fractional deviations are applied to all terms at a given post-Newtonian order.
One can convert between these two parameterizations \textit{post hoc} by requiring that the total phase correction be the same with either choice; the results shown in Figs.~\ref{fig:violinplots} and~\ref{fig:pnbounds} correspond to the parameterization used by \texttt{PhenomPNRT}.
Nevertheless, the different treatment of the spin terms may still explain the discrepancy seen at $1.5$PN, where spin effects first enter.\footnote{In the \texttt{SEOBNRT} parametrization, the PDF for $\delta \hat{\varphi}_7$ touches the prior bounds. After mapping to the \texttt{PhenomPNRT} parameterization, these tails of the distribution are down-weighted, so our final results are a good approximation to the complete PDF.}
Either parameterization offers a reasonable phenomenological description of deviations from GR; the generally close correspondence at most PN orders between results from the two models indicates that the quantities measured can be interpreted in similar ways.
For more details on each waveform model we use, see Table I of~\cite{Abbott:2018wiz}.

The long inspiral observed in GW170817 (relative to previous binary black hole signals) allows us to place the first stringent constraints on $\delta \hat \varphi_{-2}$.
This parameter is of particular interest due to its association with dipole radiation, i.e. radiation sourced by a time-varying dipole moment of the binary.
Dipole radiation is forbidden in pure GR; however, adding other long-range fields to theory---either in the gravitational sector (e.g. massless scalar-tensor theories) or non-gravitational sector (e.g. electromagnetism)---enables this new dissipative channel.
The additional energy flux induces a negative $-1$PN order correction to the phase evolution, provided that dipole radiation only contributes a small correction to the total flux predicted in GR.
The precise nature of the additional long-range fields determines the dependence of this $-1$PN correction on the various other parameters describing the binary (e.g. masses, spins, etc); in line with the theory-agnostic approach pursued here, we assume no \textit{a priori} correlation between dipole radiation and the other binary parameters by using a uniform prior on  $\delta \hat \varphi_{-2}$.

Writing the total energy flux as ${\mathcal{F}_\text{GW} = \mathcal{F}_\text{GR}(1+B c^2/v^2)}$, the leading-order modification to the phase due to theory-agnostic effects of dipole radiation is given by $\delta \hat \varphi_{-2} = -4 B/ 7$~\cite{Arun:2012hf,Barausse:2016eii}.
Combining the PDFs shown in Fig.~\ref{fig:violinplots} obtained with the \texttt{PhenomPNRT} and \texttt{SEOBNRT} waveforms, converting to a PDF on $B$ using the previous relation, and restricting to the physical parameter space $B \geq 0$ corresponding to positive outgoing flux, the presence of dipole radiation in GW170817 can be constrained to $B\leq 1.2\times 10^{-5}$.
For comparison, precise timing of radio pulses from the double pulsar PSR J0737-3039 offers some of the best current theory-agnostic constraints $|B|\lesssim 10^{-7}$~\cite{Lyne:2004cj,Kramer:2006nb,Yunes:2010qb, Barausse:2016eii,Kramer:2016kwa}. \footnote{Neutron star-white dwarf binaries offer stronger constraints than the double pulsar on certain \emph{specific} alternative theories of gravity~\cite{Freire:2012mg,Shao:2017gwu,Anderson:2019eay}, but provide comparable theory-agnostic constraints. }
This much stronger constraint arises, in part, because of the much longer observation time over which the inspirals of binary pulsars are tracked.

Though our bound on the dipole parameter $B$ is weaker than existing constraints, it is the first that comes directly from the nonlinear and dynamical regime of gravity achieved during compact binary coalescences.
In this regard, we note that for general scalar-tensor theories there are regions of parameter space where constraints from both Solar System and binary pulsar observations are satisfied, and yet new effects not present in GR appear in the frequency range of GW detectors, such as spontaneous scalarization~\cite{Damour:1993hw, Ramazanoglu:2016kul} or resonant excitation~\cite{Cardoso:2011xi, Sampson:2013jpa} of a massive field, or dynamical scalarization~\cite{Barausse:2012da, Shibata:2013pra, Palenzuela:2013hsa, Sampson:2014qqa, Shao:2017gwu}.

\section{Constraints from gravitational wave propagation}
The propagation of GWs may differ in theories beyond GR, and the deviations depend on the 
distance that the GWs travel. The search for such 
deviations provides unique tests of relativity, particularly when the distance inferred 
through GWs can be compared with an accurate, independent distance measurement from EM observations.
In GR, GWs propagate non-dispersively at the speed of light with an amplitude inversely proportional to the distance travelled.
Using GW170817, we carry out two different types of analyses to study the propagation of GWs, looking for possible deviations from GR's predictions.
The first method implements a generic modification to the GWs dispersion
relation, adding terms that correct for a massive graviton, and momentum dependent dispersion that could be apparent in Lorentz violating models~\cite{Will:1997bb,Mirshekari:2011yq}.
The second modifies the distance relation GWs follow in GR by adding correcting factors accounting for the GW's {\it gravitational leakage} into the large extra dimensions of higher-dimensional theories of gravity~\cite{Deffayet:2007kf, Pardo:2018ipy}.

\subsection{Constraints on Modified Dispersion}

In GR, gravitational waves propagate at the speed of light and are non-dispersive, leading to a dispersion relation $E^2=p^2\,c^2$.
An alternative theory may generically modify this as
 $E^2=p^2c^2+A\,p^{\alpha}c^{\alpha}$, where $A$ is the
coefficient of modified dispersion corresponding to the exponent denoted
by $\alpha$~\cite{Will:1997bb,Mirshekari:2011yq}. When $\alpha=0$,
a modification with $A>0$ may be interpreted as due to a non-zero graviton mass  
($A=m_g^2\,c^4$)~\cite{Mirshekari:2011yq}. 
It can be shown that such modified dispersion relations would lead to corrections to the GW phasing, thereby allowing us to constrain any dispersion of  GWs~\cite{Mirshekari:2011yq}.
This method, implemented in a Bayesian framework, placed bounds on $A$ corresponding to different $\alpha$  using binary black hole detections~\cite{Abbott:2017vtc}.
We apply the above method to constrain dispersion of GWs in the case of the binary neutron star merger GW170817~\cite{TheLIGOScientific:2017qsa}.
We find that GW170817 places weaker bounds on dispersion of GWs than the binary black holes.
For instance, the bound on the graviton mass $m_g$ we obtain from GW170817 is 
$9.51 \times 10^{-22}~\rm{eV}/c^2$, which is weaker compared to the bounds reported in 
\cite{Abbott:2017vtc}.
This is not surprising as GW170817 is the closest source detected so
far, and for the same SNR propagation-based tests such as this are more effective when the
sources are farther away.
This method complements the bounds on non-dispersive standard model
extension coefficients~\cite{Kostelecky:2008ts} reported in
\cite{Monitor:2017mdv} from GW170817.

\subsection{Constraints on the Number of Spacetime Dimensions}

\begin{figure}
\includegraphics[width=0.9\columnwidth]{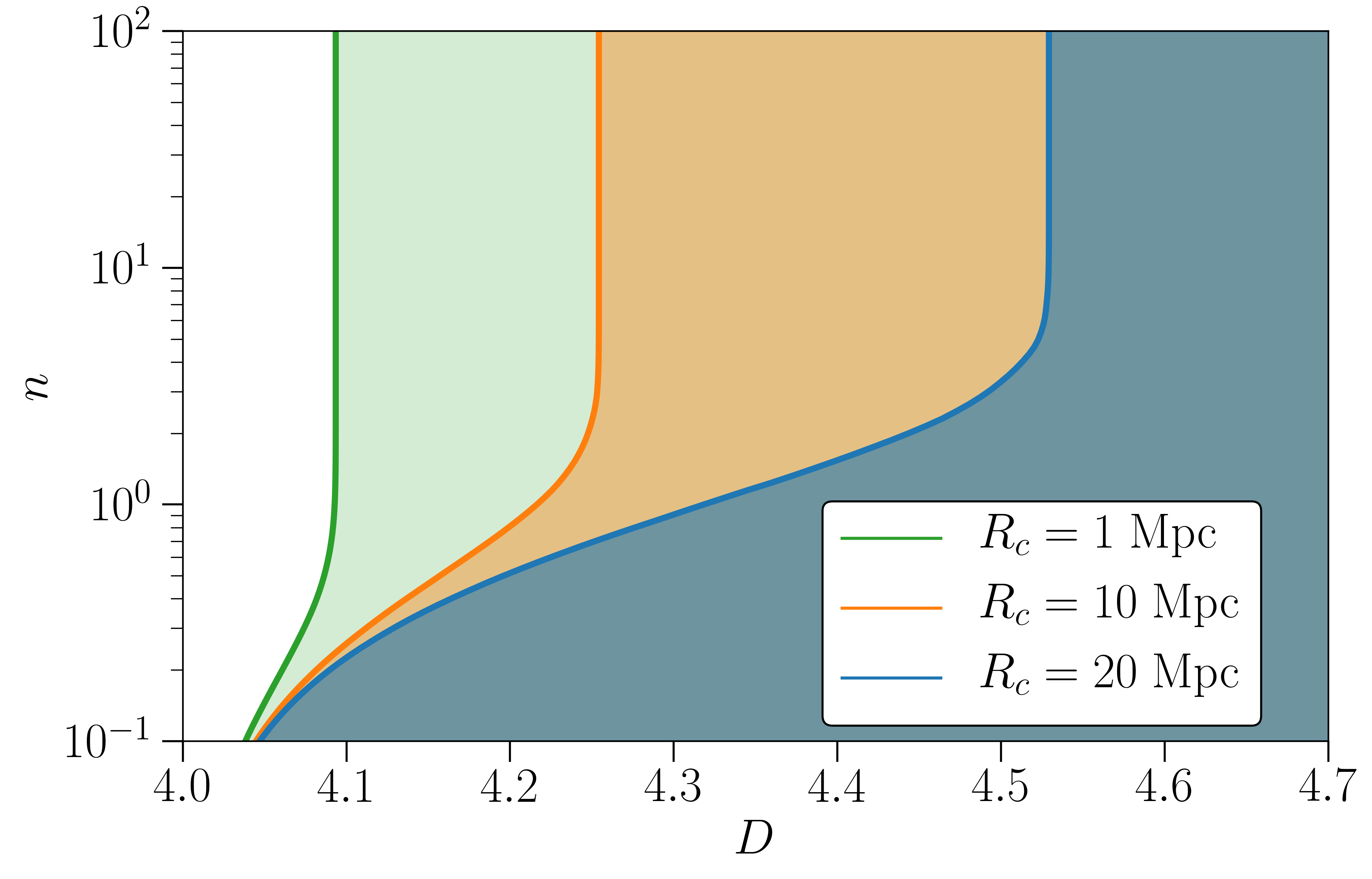}
\caption{$90\%$ upper bounds on the number of spacetime dimensions $D$, assuming fixed transition steepness $n$ and distance scale $R_c$. Shading indicates the regions of parameter space excluded by the data.}
\label{fig:dimensions}
\end{figure}

In higher-dimensional theories of gravity the scaling between the GW strain and the luminosity distance of the source is expected to be modified, suggesting a damping of the waveform due to gravitational leakage into large extra dimensions.
This deviation from the GR scaling $h_{\rm GR} \propto d_{L}^{-1}$ depends on the number of dimensions $D > 4$ and would result in a systematic overestimation of the source luminosity distance inferred from GW observations \cite{Deffayet:2007kf,Pardo:2018ipy}.
A comparison of distance measurements from GW and EM observations of GW170817 allows us to constrain the presence of large additional spacetime dimensions.
We assume, as is the case in many extra-dimensional models, that light and matter propagate in four spacetime dimensions only, thus allowing us to infer the EM luminosity distance $d_L^{\rm EM}$.
In the absence of a complete, unique GW model in higher-dimensional gravity, 
we use a phenomenological ansatz for the GW amplitude scaling and neglect all other effects of 
modified gravity in the GW phase and amplitude. 
This approach requires that gravity be asymptotically GR in the strong-field regime, while modifications due to leakage into extra dimensions start to appear at large distances from the source. 
We therefore consider gravity modifications with a screening mechanism, i.e., a phenomenological model with a characteristic length scale $R_c$ beyond which the propagating GWs start to leak into higher dimensions.
In this model, the strain scales as
\begin{equation}
\label{eq:baselinemodel}
h \propto \frac{1}{d_L^{\rm GW}} = \frac{1}{d_L^{\rm EM}}\left[1+\left(\frac{d_L^{\rm EM}}{R_c}\right)^n\right]^{-(D-4)/(2n)}
\end{equation}
where $D$ denotes the number of spacetime dimensions, and where $R_c$ and $n$ are the distance scale of the screening and the transition steepness, respectively. 
Eq.~(\ref{eq:baselinemodel}) reduces to the standard GR scaling at distances much shorter than $R_c$, and the model is consistent with tests of GR performed in the Solar System or with binary pulsars. 
Unlike the scaling relation considered in \cite{Deffayet:2007kf,Pardo:2018ipy}, notice that Eq.~(\ref{eq:baselinemodel}) reduces to the GR limit for $D=4$ spacetime dimensions.
An independent measurement of the source luminosity distance from EM observations of GW170817 allows us to infer the number of spacetime dimensions from a comparison of the GW and EM distance estimates, for given values of model parameters $R_c$ and $n$.
Constraints on the number of spacetime dimensions are derived in a framework of Bayesian analysis, from the joint posterior probability for $D$, $d_{L}^{\rm GW}$ and $d_{L}^{\rm EM}$, given the two statistically independent measurements of EM data $x_{\rm EM}$ and GW data $x_{\rm GW}$. The posterior for $D$ is then given by:
\begin{widetext}
\begin{equation}
\label{eq:probdimensions}
p(D|x_{\rm GW},x_{\rm EM}) = \int p(d_L^{\rm GW}|x_{\rm GW}) p(d_L^{\rm EM}|x_{\rm EM}) \delta(D-D(d_L^{\rm GW},d_L^{\rm EM},R_c,n))\, {\rm d}d_L^{\rm GW} {\rm d}d_L^{\rm EM}\,.
\end{equation}
\end{widetext}
As in \cite{Abbott:2018wiz}, we use a measurement of the surface brightness fluctuation distance to the host galaxy NGC 4993 from \cite{Cantiello:2018ffy} to constrain the EM distance, assuming a Gaussian distribution for the posterior probability $p(d_L^{\rm EM}|x_{\rm EM})$, with the mean value and standard deviation given by $40.7\pm 2.4$ Mpc \cite{Cantiello:2018ffy}.
Contrary to \cite{Pardo:2018ipy}, our analysis relies on a direct measurement of $d_L^{\rm EM}$ and is independent of prior information on $H_0$ or any other cosmological parameter. 
For the measurement of the GW distance, the posterior distribution $p(d_{L}^{\rm GW}|x_{\rm GW})$ was inferred from the GW data assuming general relativity and fixing the sky position to the optical counterpart while marginalizing over all other waveform parameters~\cite{Abbott:2018wiz}.
Our analysis imposes a prior on the GW luminosity distance that is consistent with 
a four-dimensional Universe, but we have checked that other reasonable prior choices do not 
significantly modify the results. 
We invert the scaling relation in Eq.~(\ref{eq:baselinemodel}) to compute 
$D(d_L^{\rm GW},d_L^{\rm EM},R_c,n)$ in Eq.~(\ref{eq:probdimensions}). 
Fig.~\ref{fig:dimensions} shows the $90\%$ upper bounds on the number of dimensions $D$, for theories with a certain transition steepness $n$ and distance scale $R_c$. Shading indicates the excluded regions of parameter space. 
Our results are consistent with the GR prediction of $D=4$.

\begin{figure}
\includegraphics[width=0.9\columnwidth]{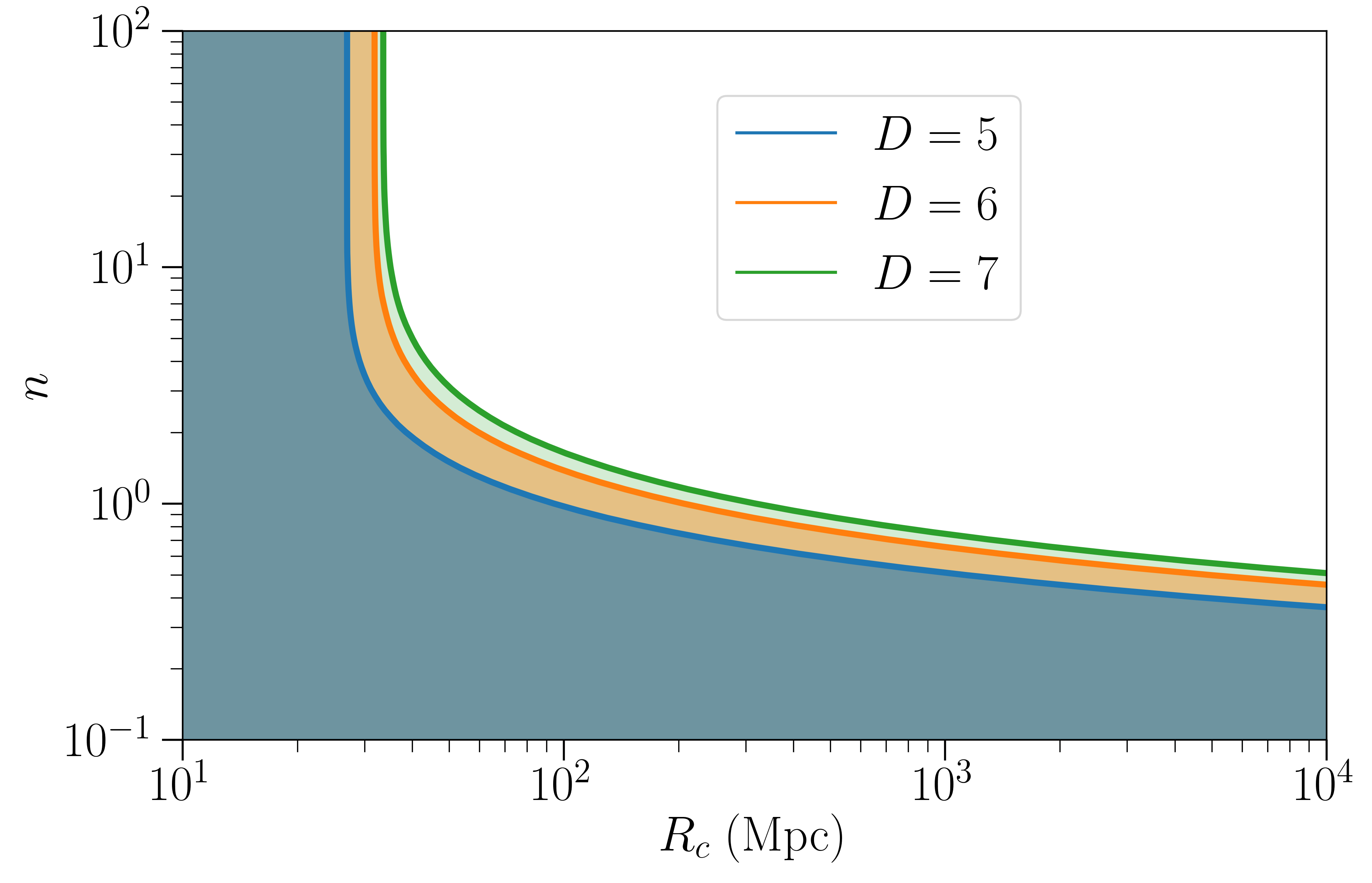}
\caption{$10\%$ lower limits on the distance scale $R_c$ (in Mpc), assuming fixed transition steepness $n$ and number of spacetime dimensions $D$. Shading indicates the regions of parameter space excluded by the data.}
\label{fig:screening}
\end{figure}

Additionally, the data allows us to infer constraints on the characteristic distance scale $R_c$ of higher-dimensional theories 
with a screening mechanism, while fixing $D$ to 5, 6 or 7.
The posterior for $p(R_c|x_{\rm GW},x_{\rm EM})$ is obtained from the joint posterior probability of $R_c, d_L^{GW}$ and $d_L^{EM}$, given by a function that is formally the same as Eq.~(\ref{eq:probdimensions}), but with $D$ and $R_c$ switching places. We fix the model parameters $D$ and $n$ and compute $R_c(d_L^{\rm GW},d_L^{\rm EM},D,n)$ by inverting the scaling relation in Eq.~(\ref{eq:baselinemodel}). 
Since we consider higher-dimensional models that allow only for a relative damping of the GW signal, we select posterior samples with $d_{L}^{\rm GW} > d_{L}^{\rm EM}$, leading to an additional step function $\theta(d_L^{\rm GW}-d_L^{\rm EM})$ in $p(R_c|x_{\rm GW},x_{\rm EM})$.
In Fig.~\ref{fig:screening}, we show  $10\%$ lower bounds on the screening radius $R_c$, for theories with a certain fixed transition steepness $n$ and number of dimensions $D>4$. Shading indicates the excluded regions of parameter space.
For higher-dimensional theories of gravity with a characteristic length scale $R_c$ of the order of the Hubble radius $R_H \sim 4\,{\rm Gpc}$, such as the well known Dvali-Gabadadze-Porrati (DGP) models of dark energy \cite{Dvali:2000hr,Lue:2005ya}, small transition steepnesses ($n \sim \mathcal{O}(0.1)$) are excluded by the data.
Our analysis cannot conclusively rule out DGP models that provide a sufficiently steep transition ($n>1$) between GR and the onset of gravitational leakage. 
Future LIGO-Virgo observations of binary neutron star mergers, especially at higher redshifts, have the potential to place stronger constraints on higher-dimensional gravity.

\section{Constraints on the Polarization of Gravitational Waves}
Generic metric theories of gravity predict up to six polarization modes for metric perturbations: two tensor (helicity $\pm 2$), two vector (helicity $\pm 1$), and two scalar (helicity $0$) modes~\cite{Eardley:1974nw,Will:2014kxa}.
GWs in GR, however, have only the two tensor modes regardless of the source properties; any detection of a non-tensor mode would be unambiguous indication of physics beyond GR.
The GW strain measured by a detector can be written in general as $h(t) = F^A h_A$, where $h_A$ are the 6 independent polarization modes and $F^A$ represent the detector responses to the different modes $A=(+,\times,{\rm x},{\rm y},{\rm b},{\rm l})$.
The antenna response functions depend only on the detector orientation and GW helicity, i.e.~they are independent of the intrinsic properties of the source.
We can therefore place bounds on the polarization content of GW170817 by studying which combination of response functions is consistent with the signal observed~\cite{PhysRevD.86.022004,PhysRevD.96.042001,Callister:2017ocg,Abbott:2018utx,Abbott:2017tlp}.

The first test on the polarization of GWs was performed for GW150914~\cite{TheLIGOScientific:2016src}.
The number of GR polarization modes expected was equal to the number of detectors in the network that observed GW150914, rendering this test inconclusive.
The addition of Virgo to the network of GW detectors allowed for the first informative test of polarization for GW170814~\cite{Abbott:2017oio}.
This analysis established that the GW data was better described by pure
tensor modes than pure vector or pure scalar modes, with Bayes factors in favor of tensor modes of more than 200 and 1000 respectively.

We here carry out a test similar to~\cite{Abbott:2017oio} by performing a coherent Bayesian analysis of the 
signal properties with the three interferometer outputs, using either the tensor or the vector or 
the scalar response functions. (Note that although the SNR in Virgo was significantly lower than in 
the two LIGO detectors, the Virgo data stream still carries information about the signal.)
We assume that the phase evolution of the GW can be described by GR templates, but the polarization content can vary \cite{Isi:2017fbj}.
The phase evolution is modeled with the GR waveform model {\tt IMRPhenomPv2} and the analysis is carried out with {\tt LALInference}~\cite{Veitch:2014wba}. 
Tidal effects are not included in this waveform model, but this is not 
expected to affect the results presented below, since the polarization test is sensitive 
to the antenna pattern functions of the detectors and not the phase evolution of the signal, as argued above. 
The analysis described above tests for the presence of pure tensor, vector, or scalar modes. We leave the analysis of mixed-mode content to future work.

If the sky location of GW170817 is constrained to NGC 4993, we find overwhelming evidence in favor of 
pure tensor polarization modes in comparison to pure vector and pure scalar modes with a (base ten) 
logarithm of the Bayes factor of $+20.81 \pm 0.08$ and $+23.09 \pm 0.08$ respectively.
This result is many order of magnitudes stronger than the GW170814 case both due to the sky position of 
GW170817 relative to the detectors and the fact that the sky position is determined precisely 
by electromagnetic observations.
Indeed if the sky location is unconstrained we find evidence against scalar modes with 
$+5.84 \pm 0.09$, while the test is inconclusive for vector modes with $+0.72 \pm 0.09$.
\section{Conclusions}
Using the binary neutron star coalescence signal GW170817, and in some cases also its associated electromagnetic counterpart, we have subjected general relativity to a range of tests related to the dynamics of the source (putting bounds on deviations of PN coefficients), the propagation of gravitational waves (constraining local Lorentz invariance violations, as well as large extra dimensions), and the polarization content of gravitational waves.
In all cases we find agreement with the predictions of GR.

The upcoming observing runs of the LIGO and Virgo detectors are expected to result in more detections of binary neutron star coalescences~\cite{Aasi:2013wya}.
Along with electromagnetic observations, combining information from gravitational wave events (including binary black hole mergers) will lead to increasingly more stringent constraints on deviations from general relativity 
\cite{Li:2011cg,Meidam:2017dgf}, or conceivably potential evidence of the theory's
shortcomings.
\section{Acknowledgements}
The authors gratefully acknowledge the support of the United States
National Science Foundation (NSF) for the construction and operation of the
LIGO Laboratory and Advanced LIGO as well as the Science and Technology Facilities Council (STFC) of the
United Kingdom, the Max-Planck-Society (MPS), and the State of
Niedersachsen/Germany for support of the construction of Advanced LIGO 
and construction and operation of the GEO600 detector. 
Additional support for Advanced LIGO was provided by the Australian Research Council.
The authors gratefully acknowledge the Italian Istituto Nazionale di Fisica Nucleare (INFN),  
the French Centre National de la Recherche Scientifique (CNRS) and
the Foundation for Fundamental Research on Matter supported by the Netherlands Organisation for Scientific Research, 
for the construction and operation of the Virgo detector
and the creation and support  of the EGO consortium. 
The authors also gratefully acknowledge research support from these agencies as well as by 
the Council of Scientific and Industrial Research of India, 
the Department of Science and Technology, India,
the Science \& Engineering Research Board (SERB), India,
the Ministry of Human Resource Development, India,
the Spanish  Agencia Estatal de Investigaci\'on,
the Vicepresid\`encia i Conselleria d'Innovaci\'o, Recerca i Turisme and the Conselleria d'Educaci\'o i Universitat del Govern de les Illes Balears,
the Conselleria d'Educaci\'o, Investigaci\'o, Cultura i Esport de la Generalitat Valenciana,
the National Science Centre of Poland,
the Swiss National Science Foundation (SNSF),
the Russian Foundation for Basic Research, 
the Russian Science Foundation,
the European Commission,
the European Regional Development Funds (ERDF),
the Royal Society, 
the Scottish Funding Council, 
the Scottish Universities Physics Alliance, 
the Hungarian Scientific Research Fund (OTKA),
the Lyon Institute of Origins (LIO),
the Paris \^{I}le-de-France Region, 
the National Research, Development and Innovation Office Hungary (NKFIH), 
the National Research Foundation of Korea,
Industry Canada and the Province of Ontario through the Ministry of Economic Development and Innovation, 
the Natural Science and Engineering Research Council Canada,
the Canadian Institute for Advanced Research,
the Brazilian Ministry of Science, Technology, Innovations, and Communications,
the International Center for Theoretical Physics South American Institute for Fundamental Research (ICTP-SAIFR), 
the Research Grants Council of Hong Kong,
the National Natural Science Foundation of China (NSFC),
the Leverhulme Trust, 
the Research Corporation, 
the Ministry of Science and Technology (MOST), Taiwan
and
the Kavli Foundation.
The authors gratefully acknowledge the support of the NSF, STFC, MPS, INFN, CNRS and the
State of Niedersachsen/Germany for provision of computational resources.

\bibliography{main}

\end{document}